\documentclass[amssymb,nobibnotes,aps,showpacs]{revtex4}
  \usepackage[dvips]{color}
  \usepackage{newlfont}
  \usepackage{graphicx}
  \usepackage{amssymb}
  \usepackage{amsmath}
  \usepackage[latin1]{inputenc}
  \usepackage{float}
  \usepackage{graphicx}
\newcommand{\br}{\begin{eqnarray}}
\newcommand{\er}{\end{eqnarray}}

\begin{document}
\title{
Two-photon interaction between trapped ions and cavity fields}
\author{F.L. Semião}
\author{A. Vidiella-Barranco}
\affiliation{Instituto de F\'\i sica ``Gleb Wataghin'' -
Universidade Estadual de Campinas, 13083-970, Campinas, S\~ao Paulo,
Brazil}
\begin{abstract}
In this paper, we generalize the ordinary two-photon Jaynes-Cummings
model (TPJCM) by considering the atom (or ion) to be trapped in a
simple harmonic well. A typical setup would be an optical cavity
containing a single ion in a Paul trap. Due to the inclusion of
atomic vibrational motion, the atom-field coupling becomes highly
nonlinear what brings out quite different behaviors for the system
dynamics when compared to the ordinary TPJCM. In particular, we
derive an effective two-photon Hamiltonian with dependence on the
number operator of the ion's center-of-mass motion. This dependence
occurs both in the cavity induced Stark-shifs and in the ion-field
coupling, and its role in the dynamics is illustrated by showing the
time evolution of the probability of occupation of the electronic
levels for simple initial preparations of the state of the system.
\end{abstract}

\maketitle

The Jaynes-Cummings model (JCM) \cite{jcm} is the most elementary
quantum model for the interaction of matter with an electromagnetic
field. In this model, a two-level atom is coupled to a single
quantized mode of a cavity field and it is one of the few fully
quantum-mechanical models that is exactly solvable. It exhibits some
unexpected nonclassical features as the revivals of the Rabi
oscillations in the atomic inversion, for instance \cite{revival}.
The quantum origin of this revival is a direct consequence of the
non-vanishing commutation relation between the creation and
annihilation operator of photons in the field mode \cite{revival}.
During the past decades, the JCM has been extended to more general
Hamiltonians including multilevel atoms \cite{levels}, multimode or
external fields \cite{modes}, multi-atom configurations
\cite{multiatom}, and multi-photon transitions \cite{mphoton}, just
to mention a few examples. All these generalizations are part of
what is called cavity quantum electrodynamics (cavity QED) and the
quantized electromagnetic field plays a fundamental role in those
settings. By the other hand, trapped ions interacting with classical
fields have gained considerable interest in the past few years,
mainly because the system dynamics closely resembles that of the
JCM, with the quantized harmonic motion of the ion's center-of-mass
playing the role of the field \cite{blockley}. Significant
experimental advances in the generation of quantum states in such a
system \cite{exp} have also indicated that trapped ions are very
suited for the study and physical implementation of quantum dynamics
\cite{review}. It seems clear that a system comprising quantum
cavity fields and trapped ions undergoing quantized harmonic motion
could bring many interesting consequences and potential
applications.  A typical setup would be that in which an ion trap is
inserted in a high finesse optical resonator so that the ion could
interact with both quantized and classical external fields
\cite{zeng}. Several authors have studied this new setting in the
framework of single-photon transitions \cite{cavion} and also Raman
transitions driven by the cavity field and a classical external
laser \cite{raman,fidio}. However, we are not restricted to
single-photon transitions in the realm of cavity QED. Other
processes, such as two-photon transitions, are particularly
important and have potential applications, e.g., in the generation
of non-classical states of the electromagnetic field, such as
squeezed states \cite{sq} or even photons with correlations that
violate classical inequalities \cite{ineq}. Nonlinear transitions in
trapped ions interacting with classical laser fields have also been
studied and applications suggested \cite{nonlinear}.

This paper is concerned with the study of atom-field two-photon
interactions when including harmonic atomic motion. We derive an
effective Hamiltonian which not only describes electronic two-photon
transitions but also contains a kind of phase-coupling between the
vibrational center-of-mass motion, the electronic degree of freedom,
and the cavity field. Another interesting feature of that
Hamiltonian is the presence of Stark-shifts depending on the number
operator of the harmonic motion. We then analyze the system dynamics
by means of the probability of finding the ion in the electronic
ground state, which is an easily accessible quantity in ion trap
experiments \cite{exp}. We discuss how this quantity is influenced
by the statistics of simple initial preparations of the quantum
state of motion. In other words, we systematically compare our model
with the important and extensively studied TPJCM without the
motional effects \cite{mphoton}.

\section{The model and results}
\begin{center}
\begin{figure}[h]
\resizebox{0.5\columnwidth}{!}{
  \includegraphics{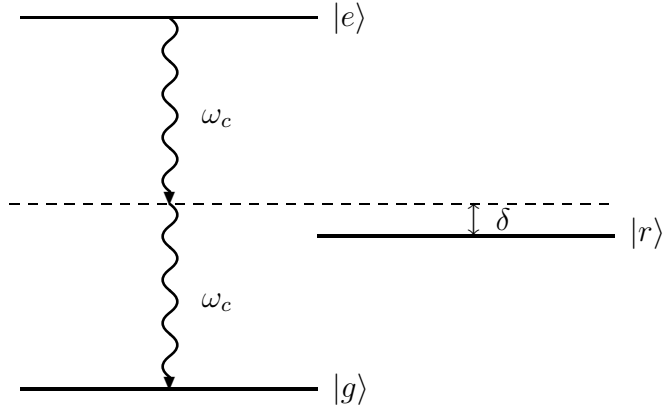}
}\caption{\label{niveis}Schematic diagram of the three-level ion
interacting with a single mode quantized field with frequency
$\omega_c$. }
\end{figure}
\end{center}

We consider a three-level ion in a cascade configuration interacting
with a single-mode quantized electromagnetic field enclosed in a
high finesse cavity. The schematic level structure is depicted in
Fig.(\ref{niveis}). It is assumed two-photon resonance between the
upper $|e\rangle$ and lower $|g\rangle$ atomic levels and the
intermediate level $|r\rangle$ is kept off-resonant. The general
Hamiltonian describing the interaction of trapped ions (or atoms)
and electromagnetic fields is discussed in several papers
\cite{blockley,zeng,cavion,raman,fidio,nonlinear}, and in our case
it reads
\begin{eqnarray}
\hat{H}=\nu \hat{a}^{\dagger}\hat{a} +
\omega\hat{b}^{\dagger}\hat{b}+E_e\hat{\sigma}_{ee}+E_r\hat{\sigma}_{rr}+
E_g\hat{\sigma}_{gg}+
[g_1(\hat{\sigma}_{gr}\hat{b}^\dagger+\hat{\sigma}_{rg}\hat{b})
+g_2(\hat{\sigma}_{re}\hat{b}^\dagger+\hat{\sigma}_{er}\hat{b})
]\cos\eta(\hat{a}^{\dagger}+\hat{a}) \label{H}
\end{eqnarray}
where $\hat{a}^{\dagger}(\hat{a})$ denotes the creation
(annihilation) operator of the center-of-mass vibrational motion of
the ion (frequency $\nu$), $\hat{b}^{\dagger}(\hat{b})$ is the
creation (annihilation) operator of photons in the field mode
(frequency $\omega$), $\hat{\sigma}_{ij}=|i\rangle\langle j|$ is a
transition atomic operator, $E_i$ is the energy of the atomic level
$|i\rangle$, $g_{1}$ e $g_{2}$ are the ion-field coupling constants
for the transitions $|g\rangle\rightarrow|r\rangle$ and
$|r\rangle\rightarrow|e\rangle$, respectively, and $\eta=2\pi
a_0/\lambda$ is the Lamb-Dicke (LD) parameter, being $a_0$ the rms
fluctuation of the ion's position in the lowest trap eigenstate, and
$\lambda$ the wavelength of the cavity field. We have taken
($\hbar=1$) and this convention will be followed in the rest of our
paper. The Lamb-Dicke parameter is the same for both transitions
(coupling constants $g_1$ and $g_2$) because both are driven by the
same (cavity) field, i.e. with the same frequency (or $\vec{k}$
vector). The cascade level structure and the field frequency [see
Fig.(\ref{niveis})] have been chosen such that by making $\delta=
E_e-E_r-\omega=E_g-E_r+\omega\gg g_1,g_2$ one can derive an
effective two-photon Hamiltonian by means of the adiabatic
elimination of the level $|r\rangle$.
The general Hamiltonian (\ref{H}) is highly nonlinear because the
function $\cos\eta(\hat{a}^{\dagger}+\hat{a})$ contains powers of
operators of the center-of-mass motion. Each term in the power
series expansion will be dependent on powers of the LD parameter
$\eta$. We can say generally that the higher the value of $\eta$ the
stronger will be the influence of nonlinear terms in the Hamiltonian
(\ref{H}). It is well know that appropriate choices of the ion-field
detuning $\delta$ may lead to different kinds of couplings in the
rotating wave approximation (RWA) \cite{blockley}. In general, it is
assumed $\delta=k\nu$, with $k$ integer, what leads to either
transitions between the trap eigenstates (k-sideband Hamiltonian,
for $k\neq 0$) or just a phase-coupling with no energy transitions
for the ion's center-of-mass harmonic motion (carrier Hamiltonian,
for $k=0$). However, it is not necessary to have $\delta=0$ in order
to forbid those transitions (make them unlikely). The same result
applies for a less demanding situation in which $\delta\ll\nu$. The
advantage of having this less demanding condition on the detuning
$\delta$ is that now it is possible to think of a situation where
the field and atom can be kept far off-resonance $(\delta\gg
g_1,g_2)$ without having excitations of the center-of-mass motion.
This may be achieved by setting $\nu\gg\delta\gg g_1,g_2$. In this
case, it is possible to obtain a carrier two-photon Hamiltonian as
we are going to show next.

The Hamiltonian (\ref{H}) in the interaction picture reads
\begin{eqnarray}
\hat{H}_I =[g_1(\hat{\sigma}_{rg}\hat{b}e^{-i\delta
t}+H.c.)+g_2(\hat{\sigma}_{er}\hat{b}e^{i\delta
t}+H.c.)]\cos\eta(\hat{a}^\dag e^{i\nu t} +\hat{a}e^{-i\nu t}),
\end{eqnarray}
which can be rewritten as
\begin{eqnarray}
\hat{H}_I=\sum_{\alpha,\beta}[g_1(\hat{\sigma}_{rg}\hat{b}e^{-i[\delta+\nu(\alpha-\beta)]
t}+H.c.)+g_2(\hat{\sigma}_{er}\hat{b}e^{i[\delta+\nu(\alpha-\beta)]
t}+H.c.)] f(\hat{a}^\dag,\hat{a};\alpha,\beta), \label{intint}
\end{eqnarray}
where
\begin{equation}
f(\hat{a}^\dag,\hat{a};\alpha,\beta)=\frac{e^{-\eta^2/2}}
{2\,\alpha!\beta!}[(i\eta)^{\alpha+\beta}+(-i\eta)^{\alpha+\beta}]a^\dag{}^\alpha
a^\beta.\label{f}
\end{equation}
Analyzing the temporal dependence of Hamiltonian (\ref{intint}), one
can see that the frequencies may be carefully chosen allowing the
RWA to be performed. In this approximation, only the slow frequency
terms are kept while the rapidly oscillating ones are discarded. For
sufficiently short interaction times, this approximation is quite
accurate as far as the coupling constants are not too strong. How
strong the coupling constants must be when compared to the other
frequencies of the problem is usually found by using time dependent
perturbation theory. In our problem, the terms in (\ref{intint})
oscillate in time as $e^{i\pm(\delta+k\nu)t}$, with $k$ integer. In
the regime $\delta\ll \nu$, the slowly oscillating terms are those
with temporal dependence $e^{\pm i\delta t}$. This only happens when
$\alpha=\beta$ in (\ref{intint}). Then, by dropping out the rapidly
oscillating terms in (\ref{intint}), we arrive at the following
approximate Hamiltonian in the original picture
\begin{eqnarray}
\hat{H}=\nu \hat{a}^{\dagger}\hat{a}+
\omega\hat{b}^{\dagger}\hat{b}+E_e\hat{\sigma}_{ee}+E_r\hat{\sigma}_{rr}+
E_g\hat{\sigma}_{gg}+f(\hat{a}^{\dagger}\hat{a})
[(g_1\hat{\sigma}_{gr}+g_2\hat{\sigma}_{re})\hat{b}^\dagger
+(g_1\hat{\sigma}_{rg}+g_2\hat{\sigma}_{er})\hat{b} ] \label{H1}
\end{eqnarray}
where,
\begin{eqnarray}
f(\hat{a}^\dag\hat{a})=e^{-\eta^2/2}\sum_{\alpha=0}^\infty\frac{(-1)^\alpha\eta^{2\alpha}
a^\dag{}^\alpha a^\alpha}{\alpha!^2}=
e^{-\eta^2/2}:J_0(2\eta\sqrt{\hat{a}^\dag\hat{a}}):,
\end{eqnarray}
and $:J_0:$ is the normally ordered zeroth order Bessel function of
the first kind. In an appendix in \cite{bana}, it is shown how to
convert $:J_0:$ in a form that does not contain the normal ordering
symbol but is an expansion in Fock basis. This is very useful for
doing the plots presented in the last part of this paper, and the
relation between both forms is in our case given by
\begin{equation}
:J_0(2\eta\sqrt{\hat{a}^\dag\hat{a}}):=\sum_{n=0}^{\infty}L_n(\eta^2)|n\rangle\langle
n|,
\end{equation}
where $L_n$ is the Laguerre polynomial. The Hamiltonian (\ref{H1})
describes a situation in which the ion's center-of-mass motion
couples to field and electronic degree of freedom in such a way that
only phases are involved, i.e. there is no real transitions between
vibrational energy levels. There is just photonic transitions taking
place. The magnitude of this phase-coupling is contained in
$f(\hat{a}^\dag\hat{a})$ which is a function of the number operator
of the vibrational motion. In the limit $\eta\rightarrow 0$ the
function $f(\hat{a}^\dag\hat{a})$ tends to the identity operator,
what corresponds to the free motion of the ion.

In order to derive an effective two-photon Hamiltonian, let us now
find the equations of motion for some relevant operators. The
starting point would be $\hat{\sigma}_{eg}$ that is supposed to be
present in the effective Hamiltonian once its function is to cause
the ion to make a direct two-photon transition from the state
$|g\rangle$ to $|e\rangle$.

The Heisenberg equation for $\hat{\sigma}_{eg}$ using (\ref{H1}) is
given by
\begin{equation}
i\frac{d}{dt}\hat{\sigma}_{eg}=(E_g-E_e)\hat{\sigma}_{eg}+
g_1f(\hat{a}^{\dagger}\hat{a})\hat{\sigma}_{er}\hat{b}^{\dagger}-
g_2f(\hat{a}^{\dagger}\hat{a})\hat{\sigma}_{rg}\hat{b}^{\dagger}
.\label{eg}
\end{equation}
The right hand side of (\ref{eg}) involves operators in the form
$f(\hat{a}^{\dagger}\hat{a})\hat{\sigma}_{ij}\hat{b}^{\dagger}$. We
need to compute the Heisenberg equations for these operators as
well, and they are given by
\begin{eqnarray}
i\frac{d}{dt}
[f(\hat{a}^{\dagger}\hat{a})\hat{\sigma}_{er}\hat{b}^{\dagger}]&=&
(E_r-E_e-\omega_c)f(\hat{a}^{\dagger}\hat{a})\hat{\sigma}_{er}\hat{b}^{\dagger}
+[f(\hat{a}^{\dagger}\hat{a})]^2[g_1\hat{b}^{\dagger}\hat{b}\hat{\sigma}_{eg}
+g_2\hat{b}^\dag{}^2(\hat{\sigma}_{ee}-\hat{\sigma}_{rr})],
\label{eqm1}\\
i\frac{d}{dt}[f(\hat{a}^{\dagger}\hat{a})\hat{\sigma}_{rg}\hat{b}^{\dagger}]&=&(E_g-E_r-\omega_c)
f(\hat{a}^{\dagger}\hat{a})\hat{\sigma}_{rg} \hat{b}^{\dagger}
+[f(\hat{a}^{\dagger}\hat{a})]^2[g_1\hat{b}^\dag{}^2(\hat{\sigma}_{rr}-\hat{\sigma}_{gg})
-g_2(1+\hat{b}^{\dagger}\hat{b})\hat{\sigma}_{eg}]. \label{eqm2}
\end{eqnarray}
The adiabatic elimination of the level $|r\rangle$ follows from
(\ref{eqm1}) and (\ref{eqm2}) by considering the condition $\delta
\gg g_1,g_2$. To make this clear, it is convenient to define new
operators in an appropriate reference frame as
$\hat{\sigma}_{ij}\rightarrow\hat{\sigma}_{ij}e^{i(E_j-E_i)t}$,
$\hat{b}(t)\rightarrow\hat{b}e^{-i\omega_c t}$ e
$\hat{a}(t)\rightarrow\hat{a}e^{-i\nu t}$. In this new frame, we may
rewrite (\ref{eg}), (\ref{eqm1}) and (\ref{eqm2}) as
\begin{eqnarray}
i\frac{d}{dt}\hat{\sigma}_{eg}&=&
g_1f(\hat{a}^{\dagger}\hat{a})\hat{\sigma}_{er}\hat{b}^{\dagger}e^{i\delta
t}+
g_2f(\hat{a}^{\dagger}\hat{a})\hat{\sigma}_{rg}\hat{b}^{\dagger}e^{-i\delta
t}, \label{egr}\\
i\frac{d}{dt}
[f(\hat{a}^{\dagger}\hat{a})\hat{\sigma}_{er}\hat{b}^{\dagger}]&=&
[f(\hat{a}^{\dagger}\hat{a})]^2[g_1\hat{b}^{\dagger}\hat{b}\hat{\sigma}_{eg}
+g_2\hat{b}^\dag{}^2(\hat{\sigma}_{ee}-\hat{\sigma}_{rr})]e^{-i\delta
t},\label{eqm1r}\\
i\frac{d}{dt}[f(\hat{a}^{\dagger}\hat{a})\hat{\sigma}_{rg}\hat{b}^{\dagger}]&=&
[f(\hat{a}^{\dagger}\hat{a})]^2[g_1\hat{b}^\dag{}^2(\hat{\sigma}_{rr}-\hat{\sigma}_{gg})
-g_2(1+\hat{b}^{\dagger}\hat{b})\hat{\sigma}_{eg}]e^{i\delta t}.
\label{eqm2r}
\end{eqnarray}
We then integrate (\ref{eqm1r}) and (\ref{eqm2r}) under the
assumption that  $\delta\gg g_1,g_2$. By substituting the result of
the integrations in (\ref{egr}) and also considering that the level
$|r\rangle$ is not initially populated, we end up with the following
equation of motion in the original Heisenberg picture
\begin{eqnarray}
i\frac{d}{dt}\hat{\sigma}_{eg}=f^2(\hat{a}^{\dagger}\hat{a})\left[\frac{g_1^2}{\delta}\hat{b}^{\dagger}\hat{b}
-\frac{g_2^2}{\delta}(1+\hat{b}^{\dagger}\hat{b})\right]\hat{\sigma}_{eg}+\frac{g_1g_2}{\delta}
f^2(\hat{a}^{\dagger}\hat{a})(\hat{\sigma}_{ee}-\hat{\sigma}_{gg})\hat{b}^\dag{}^2
\end{eqnarray}
which may be obtained from the effective Hamiltonian
\begin{eqnarray}
\hat{H}=\hat{H_0}+\hat{H}_{{{Stark}}}+\hat{H}_I\label{hdf},
\end{eqnarray}
being the free part given by
\begin{equation}
\hat{H}_0=\nu\hat{a}^\dag\hat{a}+\omega_c\hat{b}^\dag\hat{b}+\omega_c(\hat{\sigma}_{ee}-\hat{\sigma}_{gg}),
\end{equation}
the Stark shifts
\begin{equation}
\hat{H}_{{{Stark}}}=\frac{g_2^2}{\delta}f^2(\hat{a}^{\dagger}\hat{a})(1+\hat{b}^\dag\hat{b})\hat{\sigma}_{ee}
+\frac{g_1^2}{\delta}f^2(\hat{a}^{\dagger}\hat{a})\hat{b}^\dag\hat{b}\,\hat{\sigma}_{gg},\label{st}
\end{equation}
and the two-photon interaction term between ion and field given by
\begin{equation}
\hat{H}_I=\frac{g_1g_2}{\delta}f^2(\hat{a}^{\dagger}\hat{a})(\hat{\sigma}_{eg}\hat{b}^2+
\hat{\sigma}_{ge}\hat{b}^\dag{}^2).\label{itif}
\end{equation}
It describes two-photon transitions between the levels $|e\rangle$ e
$|g\rangle$ with a coupling constant that depends on energy of the
center-of-mass motion via $f^2(\hat{a}^{\dagger}\hat{a})$.

An interesting feature of the Hamiltonian (\ref{hdf}) is the
dependence of the Stark shifts upon the motion, fact that is
mathematically expressed by the presence of
$f^2(\hat{a}^{\dagger}\hat{a})$ in (\ref{st}). This is the first
remarkable feature of our model that is not present in the ordinary
TPJCM. This means that the vibrational degree of freedom will have
an important influence on the dynamics of the system not only due to
the statistics of the quantum state of motion and the phase-coupling
with the rest of the system but also via the Stark-shifts. As we are
going to see below, even small variations of the Lamb-Dicke
parameter will produce significant changes in the dynamics of the
system. This makes the model here presented useful for the
investigation of quantum aspects of the light-matter interaction.

\begin{figure}[T]
\begin{center}
\includegraphics{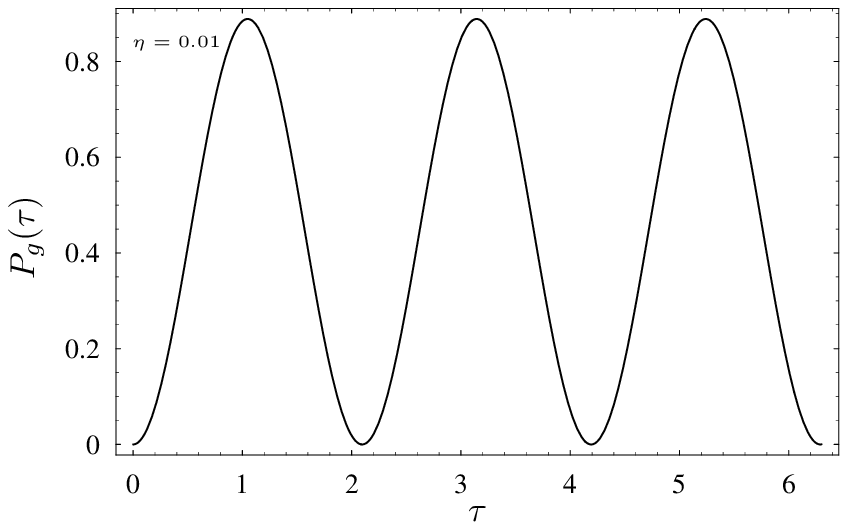}
\includegraphics{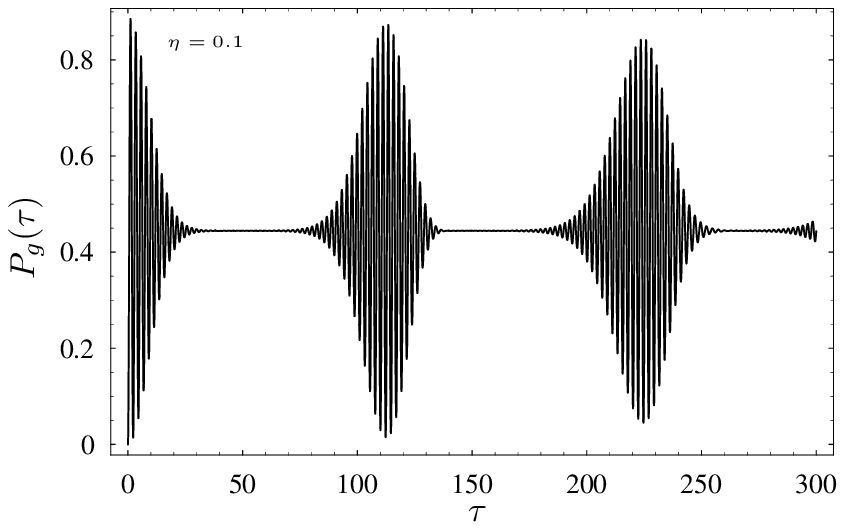}
\includegraphics{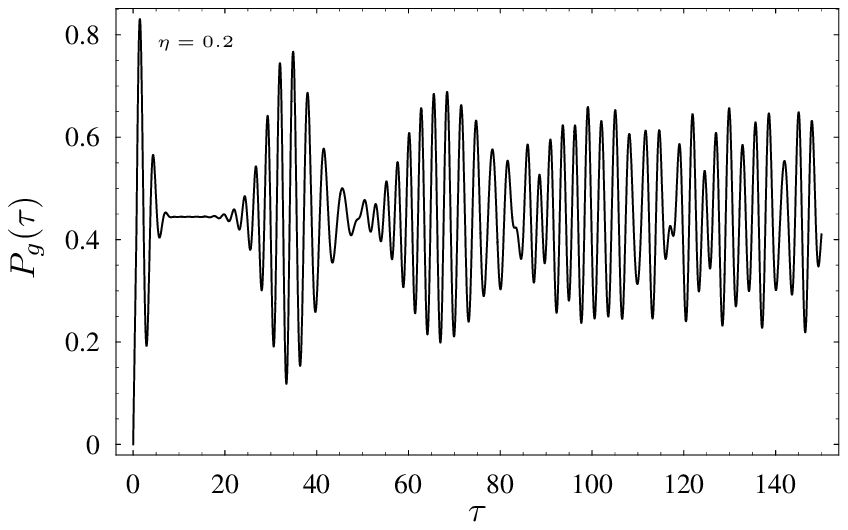}
\includegraphics{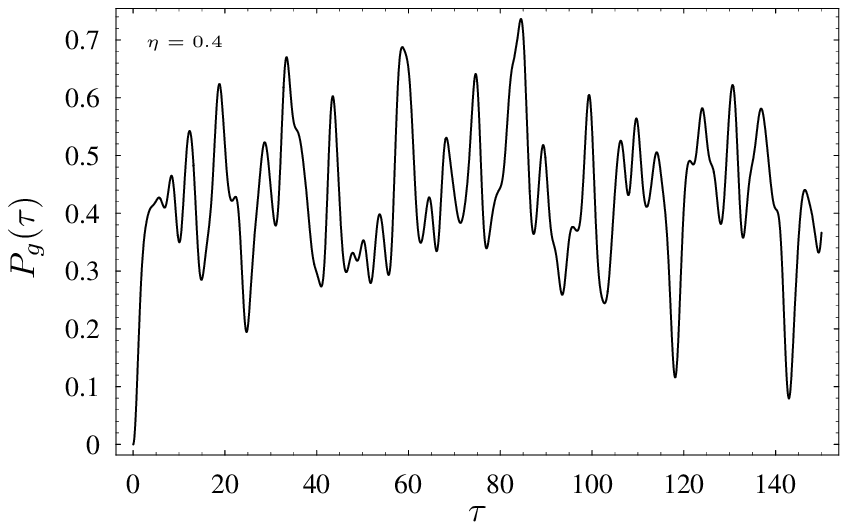}
\end{center} \caption{\label{cv}Time evolution of electronic ground state
population for different values of the Lamb-Dicke parameter. The
system has initially been prepared in the state
$|\psi(0)\rangle=|e,\alpha=2, p = 0\rangle$.}
\end{figure}

\begin{figure}[h]
\begin{center}
\includegraphics{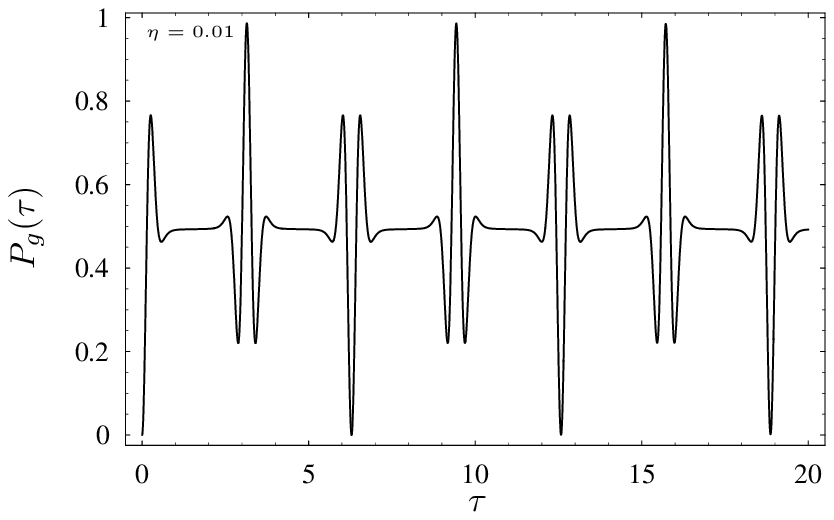}
\includegraphics{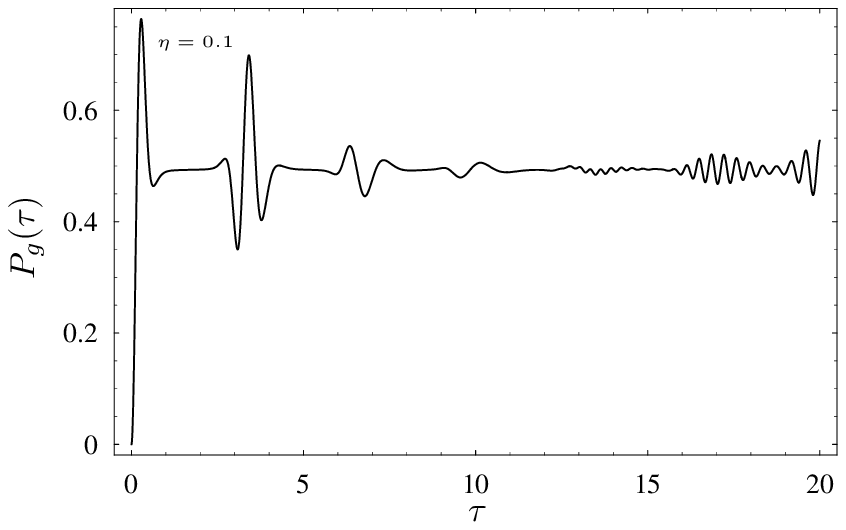}\\
\includegraphics{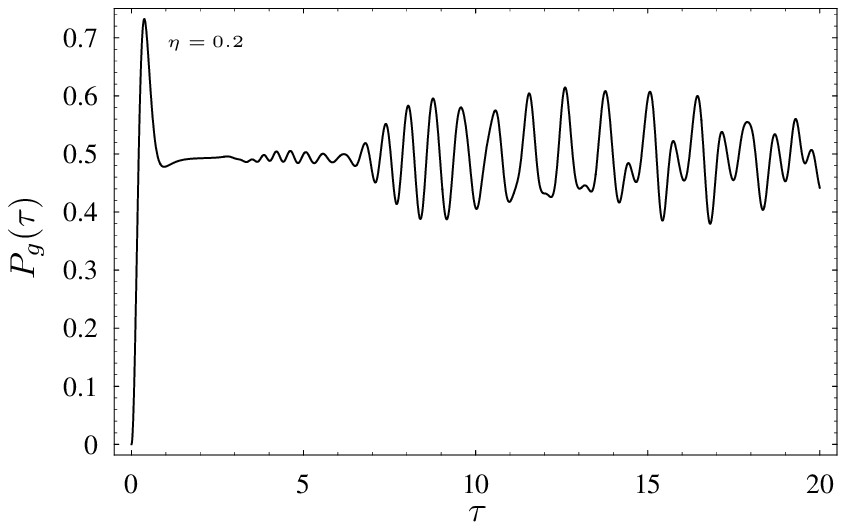}
\includegraphics{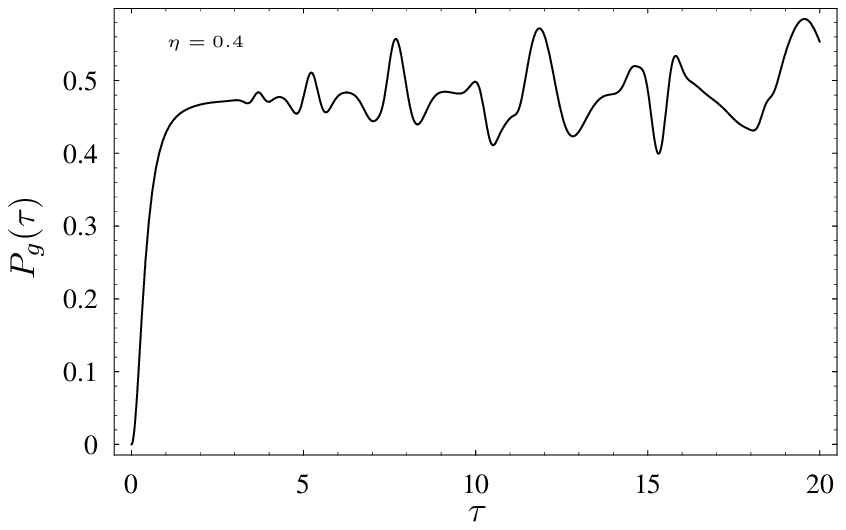}
\end{center} \caption{\label{cc}Time evolution of electronic ground state
population for different values of the Lamb-Dicke parameter. The
system has initially been prepared in the state
$|\psi(0)\rangle=|e,\alpha=2,\beta=2\rangle$.}
\end{figure}

\begin{center}
\begin{figure}[h]
\includegraphics{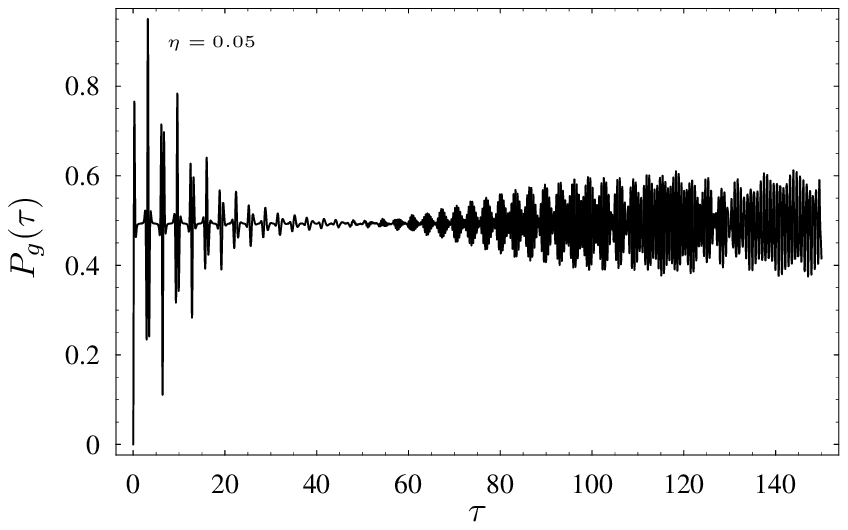}
\caption{\label{sr}Long time evolution of the electronic ground
state population. The system has initially been prepared in the
state $|\psi(0)\rangle=|e,\alpha=2,\beta=2\rangle$.}
\end{figure}
\end{center}

Once we have derived the phase-coupling two-photon Hamiltonian with
motion-dependent Stark-shifts (\ref{hdf}), which is the main result
of our paper, we now proceed with the study of the dynamics of the
electronic levels for simple but interesting and experimentally
feasible initial preparations of the system. In order to do that, we
will compute the occupation of level $|g\rangle$ that is defined as
\begin{equation}
P_g(t)=|\langle g |\psi(t)\rangle|^2.\label{pgdef}
\end{equation}
It follows from the form of (\ref{hdf}) that as long as the ion is
initially in its electronic ground state, the global state of the
system may be written as
\begin{equation}
|\psi(t)\rangle=\sum_{m=0}^\infty\sum_{n=0}^\infty
a_{mn}(t)|m,n,e\rangle+b_{mn}(t)|m,n-2,g\rangle,
\end{equation}
where the index $m$ is referred to the motion and $n$ to the field.
If one finds $a_{mn}(t)$ an $b_{mn}(t)$, the desired probability
$P_g(t)$ is just the summation in $m$ and $n$ of the squared
absolute value of $b_{mn}(t)$. In the interaction picture, the
coefficients obey the following system of differential equations
\begin{eqnarray}
i \frac{d}{dt}a_{mn}(\tau)&=&\chi_1\,a_{mn}(\tau)+\chi_2\,b_{mn}(\tau),\nonumber\\
i
\frac{d}{dt}b_{mn}(\tau)&=&\chi_3\,b_{mn}(\tau)+\chi_2\,a_{mn}(\tau),\label{s}
\end{eqnarray}
where we defined
\begin{eqnarray}
g&\equiv &\frac{g_1g_2}{\delta};\,\,\,\tau\equiv
gt;\,\,\,r=\frac{g_1}{g_2};\,\,\,\chi_1\equiv\frac{f^2(m)}{r}(n+1);\nonumber\\
\,\,\, \chi_2&\equiv
&\sqrt{(n+1)(n+2)}f^2(m);\,\,\,\chi_3=rf^2(m)(n+2).\nonumber\\
\end{eqnarray}
The solution of (\ref{s}) gives one all the information available
about the physical system described by the Hamiltonian (\ref{hdf}).
For the purposes of this paper though, we are interested in the
simplest case $g_1 = g_2$ $(r=1)$. Also, we will be particularly
interested in initial preparations given either by
\begin{eqnarray}
a_{mn}(0)&=&e^{-|\alpha|^2/2}\frac{\alpha^m}{\sqrt{m!}}\,\delta_{np}\nonumber\\
b_{mn}(0)&=&0\label{p1}
\end{eqnarray}
or
\begin{eqnarray}
a_{mn}(0)&=&e^{-(|\alpha|^2+|\beta|^2)/2}\frac{\alpha^m\beta^n}{\sqrt{m!n!}}\nonumber\\
b_{mn}(0)&=&0.\label{p2}
\end{eqnarray}
In both cases the ion is initially in the electronic excited state
$|e\rangle$, and particulary in (\ref{p1}) the motion is in the
coherent state $|\alpha\rangle$ and the field in the Fock state
$|p\rangle$ containing exactly $p$ photons, while in (\ref{p2}) the
motion and field are initially prepared in coherent states
$|\alpha\rangle$ and $|\beta\rangle$, respectively. The experimental
generation of vacuum Fock and coherent states of motion for trapped
ions have already been reported \cite{exp} as well as of
electromagnetic cavity fields \cite{coh cav}. The solution of
(\ref{s}) for $r=1$ and the ion initially in the excited state is
\begin{eqnarray}
a_{mn}(\tau)&=&a_{mn}(0)\left[\cos(\Lambda_{mn}\tau/2)+i\frac{f^2(m)}{\Lambda_{mn}}\sin(\Lambda_{mn}\tau/2)\right]
\nonumber\\
b_{mn}(\tau)&=&-2\,
i\,a_{mn}(0)\frac{\chi_2}{\Lambda_{mn}}\sin(\Lambda_{mn}\tau/2),\label{sol}
\end{eqnarray}
where $\Lambda_{mn}\equiv\sqrt{f^4(m)+4\chi_2^2}$. The sought
probability is finally found to be
\begin{equation}
P_g(\tau)=\sum_{m=0}^\infty\sum_{m=0}^\infty|a_{mn}(0)|^2
\frac{4\chi_2^2}{\Lambda_{mn}^2}\sin^2(\Lambda_{mn}\,\tau/2).\label{pg}
\end{equation}
In what follows we will be studying the effect of the motion on
$P_g(\tau)$. Mathematically, the ordinary TPJCM may be obtained by
making $\eta\rightarrow0$. We note that both the frequencies and
amplitudes of each term in equation (\ref{pg}) have distinct
contributions from the quantized field and from the ion's
vibrational motion. For the first initial preparation (\ref{p1}), we
clearly see [Fig.(\ref{cv})] how strong the influence of the atomic
motion is because slight changes in the Lamb-Dicke parameter $\eta$
lead to quite different behaviors. The expected Rabi oscillations
found in the TPJCM are modified as the parameter $\eta$ is
increased. For small values of $\eta$, i.e. in the limit of the
TPJCM, the coherent state of motion induces almost complete periodic
collapses and revivals. By increasing $\eta$, the dynamics changes
again and looses that periodicity until a complete irregular pattern
takes place. Similar strong effects of the harmonic motion on the
field dynamics in cavity QED setups have been already reported, see
for instance Di Fidio {\emph{et al}} \cite{fidio}. They consider a
trapped ion in a Raman configuration interacting with a cavity field
and an external laser. The main difference between our model and the
one treated in \cite{fidio} is that the two-photon configuration is
in essence a periodic model while the Raman coupling does not
necessarily lead to a periodic behaviour. Therefore, the inclusion
of the atomic motion might not produce the same effects on those two
different cavity-QED setups. In fact, the influence of the atomic
motion should become more evident in the model treated here, as the
oscillations induced by the atomic motion are in general not
periodic, in contrast to the oscillations due to the interaction
with the field. Considering now the second initial preparation
(\ref{p2}), for an initial coherent state for the field, the
dynamics is also modified by the harmonic motion [Fig.(\ref{cc})].
In this case, the characteristic (almost) periodic evolution found
in the in the ordinary TPJCM and reported in several papers
\cite{mphoton}, is again modified by the increasing of $\eta$. The
beats due to the statistics of the initial state of the
center-of-mass motion (coherent state) clearly destroy the regular
patterns. Moreover, for this initial preparation in which both the
field and the center-of-mass motion of the ion are prepared in
coherent states, we found for intermediate values of the LD
parameter the interesting behavior of revivals ocurring at longer
times \cite{super}, as shown in Fig.(\ref{sr}). This \emph{super
revival} or revival of revivals is a revival at long times of the
Rabi oscillations and the ordinary short-time revivals. That is
another special feature happening for this model that is not present
in the TPJCM. Once the dynamics in the TPJCM is almost periodic for
short and long interaction times, the concept of super revivals has
no meaning in this case. All these different effects indicate some
of the several interesting processes that might arise when
considering cavity quantum electrodynamics with trapped ions,
specially the model proposed here. We note, from the examples above,
that the effects of the quantized field and the atomic motion are
somehow superimposed and have peculiar features, such as, for
instance, a dependence of the Stark shifts terms on the atomic
motion.

\section{Conclusions}
We have investigated the interaction of a trapped ion with the
quantized cavity field via two photon transitions. Particularly in
this paper, we have studied this system focusing on the fundamental
aspects of the understanding of light-matter interaction. We have
derived an effective Hamiltonian under the rotating wave
approximation and presented its analytical solution. This effective
Hamiltonian contains motion-dependent Stark-shifts and a coupling
constant which is a function of the intensity of motion, i.e. it is
a function of the massive oscillator number operator. Both features
are not present in the original two-photon Hamiltonian of the TPJCM.
In particular, we have calculated the evolution of the population of
the electronic levels, and found how the center-of-mass motion
decisively changes the dynamics. In general, a treatment of the open
system by including cavity losses, spontaneous electronic emission,
and so forth, is necessary, although it rarely possesses a closed
analytical solution. Some important and exceptional cases that admit
analytical solutions for especial regimes were pointed out and
treated by Di Fidio {\emph{et al}} \cite{fidio}. Even though the
system considered in \cite{fidio} is not identical to ours as they
treat the case of Raman transitions, most their findings may be
applied to our problem. As they point out, the inclusion of these
realistic conditions will certainly have a destructive effect as
long as quantum coherences are involved. In practise, some behaviors
of $P_g(\tau)$ presented here will not be observed in the current
experimental setups, and it is the case for the super revival
phenomenon. However, it is always important to investigate the
presence or not of these typical quantum effects. On the other hand,
differences between our model and the ordinary TPJCM arise also at
shorter times as, indicated in Fig.{\ref{cv}} and Fig.{\ref{cc}}.
Therefore, it would be possible to observe some of those effects
with the achievement of the strong coupling regime; although losses
and dephasing effects are present, they will have a much weaker
magnitude than the cavity-ion coupling itself. The system comprising
cavities and trapped ions studied here is now under intense
experimental investigation with significant control improvements
\cite{blatt}, and efforts are being made in order to reach the
strong coupling regime.

This work is partially supported by CNPq (Conselho Nacional para o
Desenvolvimento Cient\'\i fico e Tecnol\'ogico), and FAPESP (Funda\c
c\~ao de Amparo \`a Pesquisa do Estado de S\~ao Paulo) grant number
02/02715-2, Brazil.


\end{document}